\documentstyle{l-aa}

\begin{document}

\thesaurus{12 (12.03.1; 12.03.3; 12.03.4; 12.12.1)} 

\title{Cosmology with Galaxy Clusters: I. Mass Measurements and Cosmological
Parameters}

\author{Asantha R. Cooray}
%\offprints{Asantha R. Cooray}

\institute{Department of Astronomy and Astrophysics, University of Chicago, Chicago IL 60637, USA. E-mail: asante@hyde.uchicago.edu}

\date{Received:  March 1, 1998; accepted }
\maketitle
%------------------------------------------------------------------------------

\begin{abstract}

Under the assumption that the gas mass fraction of galaxy clusters
estimated out to an outer hydrostatic radius is 
constant, it is possible to constrain the
cosmological parameters by using the
angular diameter distance relation with redshift.
We applied this to a sample of galaxy clusters  
from redshifts of 0.1
to 0.94, for which
published gas and total masses are available from X-ray data.
After scaling the gas fraction values to the $r_{500}$ radius (Evrard, Metzler,
\& Navarro 1996), 
we find an apparent decrease in gas fractions at high redshifts,
which can be
scaled back to the mean gas fraction value at z $\sim$ 0.1 to 0.2 of
(0.060 $\pm$ 0.002) $h^{-3/2}$, when 
$\Omega_m = 0.55^{+0.35}_{-0.23}$ ($\Omega_m + \Omega_{\Lambda} =1$, 
1-$\sigma$ statistical error).
However, various sources of systematic errors can contribute to the
change in gas mass fraction from one cluster to another, and
we discuss such potential problems in this method.
%Given the selection biases that may go in to the present
%data, results from a  well defined sample of galaxy clusters,
%such as that might be available with AXAF in X-ray and
%with interferometers in SZ, should allow
%strong constraints on the cosmological parameters in the near future.

\end{abstract}
%------------------------------------------------------------------------------
% User-supplied List of keywords.


%------------------------------------------------------------------------------
\section{Introduction}
In recent years, measurements of gas mass fraction in galaxy clusters together
with the universal value for the cosmological baryon density as derived from
nucleosynthesis arguments have been used to constrain
the cosmological mass density of the universe. There are two basic
assumptions used in such an analysis: (1) that the galaxy
clusters are the largest virialized systems in the universe
and based on hierarchical clustering models that clusters
represent composition of the universe as whole, and
(2) that the gas mass fraction when measured out to
a standard (hydrostatic) radius is constant.
Evrard (1997) applied this argument to a sample of galaxy clusters
studied by David, Jones \& Forman (1995) and White \& Fabian (1995), with
redshifts up to 0.2. The gas mass fraction values were
scaled to the $r_{500}$ radius first defined by
Evrard, Metzler \& Navarro (1996),
which has shown to be a conservative estimate of the
outer hydrostatic radius
based on numerical simulations. 
When both gas and total (virial) masses are estimated
out to this radius, then the assumption that all clusters
should have the same gas mass fraction is well justified.
The mean gas mass fraction, as derived by Evrard (1997), for the low
redshift clusters is $\bar{f}^{\rm X-ray}_{\rm gas} 
(r_{500}) = (0.060 \pm 0.003)\, 
h^{-3/2}$. Recently, Myers {\it et al.} (1997) observed the 
Sunyaev-Zel'dovich (SZ) effect  (Sunyaev \& Zel'dovich 1980) 
in a sample of low redshift
clusters, and derived a mean SZ gas mass
fraction, $\bar{f}^{\rm SZ}_{\rm gas}$,
 that ranges from (0.061 $\pm$ 0.011) $h^{-1}$ to
(0.087 $\pm$ 0.030) $h^{-1}$ (we refer the reader to Birkinshaw 1998 for
a recent review on the SZ effect). Using the mean cluster
gas fraction and the nucleosynthesis derived value for the
$\Omega_b$,
both Evrard (1997) and Myers {\it et al.} (1997)
put constraints on the cosmological mass density of
the universe, $\Omega_m$, with some dependence on the Hubble constant. 

If cluster gas fractions are indeed constant in a sample
of galaxy clusters when calculated
out to the outer hydrostatic
radius, then it is possible to
constrain the cosmological parameters based on their dependence
in the angular diameter distance relation with redshift. 
In Section 2 of this paper, we present the potential possibility
of using cluster gas fraction as a {\it standard candle} 
and apply this to a sample of $\sim$ 53 clusters that
range in redshifts from 0.1 to 0.94, for which published
data on masses are available from literature. In Section 3,
we discuss various systematics uncertainties and possible
selection effects in this method.
%In the same section, we
%study the possibility of using
%a well defined sample of galaxy clusters, as that
%would soon be available with AXAF in X-ray and
%with interferometers in SZ, to constrain the cosmological
%parameters.

\section{Method \& Results}

In general, the gas mass of a galaxy cluster is calculated
by integrating the electron number density profile, $n_{\rm e} (r)$,
over an assumed shape. The X-ray emission
data are usually fitted to an apriori assumed distribution, and presently
the $\beta$-model is used as the preferred option.
The gas mass within the galaxy cluster is given by:
\begin{equation}
M_{\rm gas}(\leq R) = m_{p} \int_{0}^{r} n_{\rm e0} \left( 1+ \frac{r'^2}{r_c^2} \right)^{-\frac{3}{2} \beta} d^3r',
\end{equation}
where $n_{e0}$ is the central number density, $r_c$ is the core
radius of the gas distribution, and $\beta$ is a parameter that describes
the shape of the distribution (e.g., Cavaliere \& Fusco-Femiano 1976). 
For the purpose of this paper, 
when $n_{\rm e0}$ is determined through the X-ray emission:
\begin{equation}
M^{\rm X-ray}_{\rm gas} (\leq \theta) \propto D_{\rm A}^{5/2},
\end{equation}
while when the SZ decrement (or increment) 
towards a cluster is used to determine $n_{\rm e0}$:
\begin{equation}
M^{\rm SZ}_{\rm gas} (\leq \theta) \propto D_{\rm A}^{2},
\end{equation}
where D$_{\rm A}$ is the angular diameter distance to the cluster.

The total virial mass of a galaxy cluster can be known through the
measurements of the intercluster electron temperature profile, $T_{\rm e}(r)$,
based on the X-ray
spectral data. Under the assumption of virial equilibrium and
assuming an isothermal temperature distribution with electron temperature of
$T_{\rm e0}$, the
total mass is:
\begin{equation}
M_{\rm tot} (\leq R) =  - \frac{k_B T_{\rm e0} r}{G \mu m_p} \frac{d \log{n_{\rm e}(r)}}{d \log{r}},
\end{equation} 
and 
\begin{equation}
M_{\rm tot} (\leq \theta) \propto D_{\rm A}.
\end{equation}
Apart from X-ray temperature measurements,
the total mass can be estimated using strong gravitational
lensing, which allows mass measurements
 near the core region of the cluster where
background sources are lensed to arcs, weak lensing and cluster internal
velocity dispersions, which allow
mass measurements to outer regions of the cluster.
However, contrary
to X-ray and optical virial analysis 
derived masses, both SZ and lensing masses measure the
mass along the cylindrical line of sight across
the cluster. The
total masses derived from X-ray temperature and strong lensing
seem to be off by factors of 2 to 3, while
weak lensing, velocity dispersion, and 
X-ray temperature masses seem to agree with each
other at very large radii.
The reasons for difference between X-ray temperature and
strong lensing masses have been studied based on the
presence of magnetic fields (Loeb \& Mao 1994; Miralda-Escud\'e \& Babul 1995) 
and cluster cooling flows (Allen 1997). 
In the present paper, we use the X-ray temperature data as
a measurement of the total (virial) mass, since most of the clusters
in our sample do not have published lensing and optical virial mass 
measurements yet.

The two gas fraction measurements based on the X-ray total mass is:
\begin{equation}
f^{\rm X-ray}_{\rm gas} \propto D_{\rm A}^{3/2},
\end{equation}
and
\begin{equation}
f^{\rm SZ}_{\rm gas} \propto D_{\rm A},
\end{equation}
for X-ray and SZ gas masses respectively.

For Friedmann-Lema$\hat{i}$tre cosmological models, the angular 
diameter distance, D$_{\rm A}$, can be written as:
\begin{equation}
D_{\rm A} = \frac{c}{H_{0} (1+z) \sqrt{\kappa}} \chi \left(x\right),
\end{equation}
where,
\begin{equation}
x = \sqrt{\kappa} \int_{0}^{\rm z}\left[ (1+z')^2(1+\Omega_M z') - z' (2+z') \Omega_{\Lambda} \right]^{-\frac{1}{2}} dz'.
\end{equation}
For $\Omega_M + \Omega_{\Lambda} =1$ 
(flat universe), $\chi(x) = x$ and $\kappa=1$, 
for  $\Omega_M + \Omega_{\Lambda} > 1$ (closed universe)
$\chi(x) = \sin(x)$ and $\kappa = 1- \Omega_M - \Omega_{\Lambda}$, 
and for $\Omega_M + \Omega_{\Lambda} < 1$ (open universe),  
$\chi(x) = \sinh(x)$ and $\kappa = 1 - \Omega_M - \Omega_{\Lambda}$. 
When $\Omega_{\Lambda}$ =0, the angular diameter distance has a well known
analytical closed form solution, while for more general
case with a cosmological constant, the integral over redshift must be
carried out (Carroll, Press \& Turner 1992).

Using published data, we compiled a sample of galaxy clusters and groups
with redshifts 0.008 to 0.94. Most of the low redshift cluster and group
data are from White, Jones \& Forman (1997) who
presented results of an X-ray image analysis of 207 clusters
based on {\it Einstein} data. These cluster range in redshift from $\sim$
0.008 to 0.4, with few clusters between $z$ $\sim$ 0.3 and 0.4.
 Our high redshift ($z> 0.4$)
cluster sample is derived from various published data and includes:
AXJ2019+1127 ($z=0.94$, Hattori 1997), MS1054-03 ($z=0.829$, 
Donahue {\it et al.} 1997), MS0451-03 ($z=0.54$, Donahue 1996), and
Cl0016+16 ($z=0.554$, Neumann \& B\"ohringer 1996). In order
to perform an unbiased test, we calculated both gas and total masses 
 in a universe with $\Omega_m = \Omega_{\Lambda} =0$
and H$_0 = 100$ $h^{-1}$ km s$^{-1}$ Mpc$^{-1}$, and
scaled them to the $r_{500}$ radius based on relations
presented by Evrard (1997):
\begin{equation}
r_{500} (T_{\rm e0}) = (1.24 \pm 0.09) \left( \frac{T_{\rm e0}}{10\, {\rm keV}} \right)^{\frac{3}{2}}\, h^{-1}\, {\rm Mpc}.
\end{equation}
At the $r_{500}$ radius, the gas fraction is given by:
\begin{equation}
f_{\rm gas} (r_{500} (T_{\rm e0})) = f_{\rm gas} (r) \left( \frac{r_{500}(T_{\rm e0})}{r}\right)^{\eta}
\end{equation}
where $\eta$ is a scaling parameter determined by numerical simulations
and $r$ is the radius out to which the gas fraction is first 
calculated using mass measurements in literature.
Evrard (1997) used $\eta=0.17$, but showed that for real data it ranges
from 0.13 to 0.17, based on which clusters are included.
We calculated our gas fraction values at $r_{500}$ using
$\eta$ values ranging from 0.13 to 0.17, but found that
the final result presented here is not heavily dependent on it.
This is primarily due to the fact that for massive clusters 
 with cluster gas temperatures close to 10 keV, 
the ratio between
r$_{500}$ and $r$ is within few percent of each other.
From our total sample of $\sim$ 215 clusters, we removed all
of the clusters with redshifts less than 0.1. This is primarily
due to the fact that the low redshift sample is mainly made
of group size clusters with electron temperatures less than 3 keV,
and the baryonic fraction of such clusters are primarily
dominated by the baryons in galaxies and stars, instead of the
intergroup gas component (Fukugita {\it et al.} 1997).

The mean baryonic fraction of galaxy clusters between
redshifts of 0.1 and 0.15 is (0.060 $\pm$ 0.002) $h^{-3/2}$, which
is in agreement with the value derived by Evrard (1997) of
(0.060 $\pm$ 0.003) $h^{-3/2}$. 
We normalized the cluster gas mass fractions to our low redshift mean gas
fraction.
In Fig.\ 1, we show the observed normalized cluster gas
fraction with redshift. For completeness, we 
show the whole sample, including
clusters and groups
with redshifts less than 0.1.
The apparent rise in the gas fraction with redshift at the low
redshifts ($z \sim$ 0.008 to 0.1)
is primarily due to a selection bias: as you go higher in
redshift more massive clusters are selected in X-ray studies which have
an increasing gas mass component. 

\begin{figure}
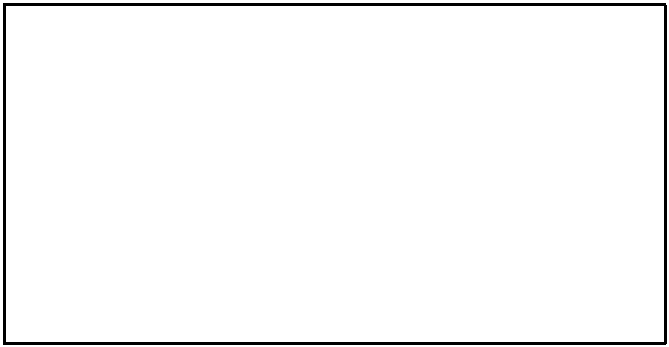

\picplace{4.5 cm}
\caption[]{The observed gas mass fraction of galaxy clusters calculated
with an angular diameter distance assuming $\Omega_m = \Omega_{\Lambda} =0$,
and normalized the low redshift mean gas mass fraction of
(0.060 $\pm$ 0.002) $h^{-3/2}$. If the universe is indeed $\Omega_m = \Omega_{\Lambda} =0$ then the observed gas fraction throughout the redshift
plane is expected to be constant. The apparent decrease in the gas mass
fraction at high redshifts can be explained with a
cosmological model where $\Omega_m$  is low. 
The drawn curve corresponds to
$\Omega_m =0.55$ ($\Omega_m+ \Omega_{\Lambda} = 1$).}
\end{figure}

As shown in Fig.\ 1, the high redshift clusters have gas fractions lower than
the low redshift sample. 
The decrease in gas fraction at high redshift can be explained in
a universe with a low $\Omega_m$. Considering a $\chi^2$ minimization
to the observed data with the angular diameter distance, we derived
a best fit $\Omega_m$ of 0.55$^{+0.35}_{-0.23}$ (1-$\sigma$ statistical
error), with $\Omega_m + \Omega_{\Lambda} =1$, at a minimum $\chi^2$ of 
62.5 for 53 clusters. 
The data are also consistent with
$\Omega_m < 0.72$ (90\% confidence interval) for an open universe
with $\Omega_{\Lambda}=0$.

\section{Discussion}

The primary assumption in constraining the cosmological parameters
using the angular diameter distance
is that the gas fraction at high redshift is same as the
fraction at low redshift clusters. It is likely
that gas fractions may vary from one cluster to another, but
numerical simulations seem to strongly suggest that
gas fractions measured to the outer hydrostatic
radius is constant and is not significantly
influenced by hydrodynamical effects
(e.g., Evrard, Metzler \& Navarro 1996; Evrard 1997;
Lubin {\it et al.} 1996).
In Shimasaku (1997), the assumption of constant gas (baryonic) fraction
was used to put constrains on $\sigma_8$, the rms linear
fluctuations on scales of 8 h$^{-1}$ Mpc, and on $n$, the
slope of the fluctuation spectrum.
However, possible reasons for
fluctuations in the gas fraction needs to be considered,
and we discuss briefly various	
systematic uncertainties and biases in this method.
These include gas clumping, presence of magnetic
fields and magnetic pressure support, various non-gravitational
processes such as heating and cooling in galaxy clusters,
additional baryonic contributions,
and a possible phase of gas injection from galaxies in clusters
to intercluster medium.

\subsection{Gas Clumping \& Magnetic Fields}
Given that the gas masses are estimated from X-ray emission, which
measures the $<n_e^2>$, one needs to assume that the gas is smoothly
distributed within the intercluster space. However, if gas is clumped 
with a clumping factor $C$ $(=<n_e^2>^{1/2}/<n_e>)$, then the amount of
gas required to produce a fixed X-ray emission will reduce by $C$.
The observational evidence against clumping is presented in Evrard (1997).
Unlike the X-ray gas mass, the gas mass derived using the SZ effect
is not subject to gas clumping as it measures $<n_e>$ directly (although the
Hubble constant that could be derived by combining SZ and X-ray data
is subject to gas clumping effects). Therefore, one should be
able to constrain clumping in galaxy clusters based on a set of
SZ and X-ray data for a cluster sample. Currently, there
are 4 mass estimates from SZ effect (Myers {\it et al.} 1997), with
a mean gas fraction of $(0.087 \pm 0.030)$ $h^{-1}$ to (0.061 $\pm$ 0.011)
$h^{-1}$. If we used the mean X-ray derived gas fraction, and under the
assumption that the difference between X-ray and SZ gas fractions
is due to a combination of
gas clumping and the Hubble constant,
an upper limit to the clumping factor $C$ is $\sim$ 1.24 when $h=0.65$.
There is also the possibility that gas clumping
is nonisothermal such that $C = <n_e^2 T_e>^{1/2}/<n_e><T_e>$, but,
 constraints on such a gas clumping model will only be possible after
reliable cluster gas temperature profiles are measured.
In future, this should be possible with planned X-ray missions
such as AXAF, XMM and Astro-E.

The presence of magnetic pressure support has been discussed as
a solution to explain the difference between X-ray virial and
strong gravitational lensing mass in the core regions of the
cluster. Given that our method uses the fact that
gas fraction out to an outer hydrostatic region is constant,
it is unlikely that magnetic fields play a major role in
varying the cluster gas fraction from one cluster to another.
However, the difference in gas fraction near the core regions
of the cluster may be explained using magnetic fields.
Here again, present data do not allow reliable estimates
on the amount of magnetic pressure support in galaxy clusters.

\subsection{Heating \& cooling in galaxy clusters}
Cluster cooling flows have been observed up to $\sim$ 40\% of the
X-ray clusters, and it is expected that $\sim$ 10\% of the observed
X-ray emission in such clusters are due to cooling flows (Fabian 1994).
Such cooling flows are expected to increase the gas fraction
near the core region of the 
cluster where such effects exist. However, the apparent increase in the
cluster gas fraction from high to low
redshift cannot be fully explained
as due to cooling flows, since the low redshift cluster sample
both contains high cooling flow and low cooling flow
clusters, with gas fractions that agree with each other,
and that there is little evidence for cooling flows
in the high redshift sample. 
If a heating process
exists within galaxy clusters then the cluster
gas fraction may be lowered. However, no physical mechanism
has yet been suggested which can lower the cluster gas
fraction.

\subsection{Additional baryonic contributions}	

The arguments used here completely ignored the presence of
baryons other then gas in clusters.
It is well known that the baryonic component of the
massive clusters are dominated by the X-ray emitting 
cluster gas rather than the 
baryons associated with stars and galaxies
in clusters. According to White {\it et al.} (1993), the mass
ratio of stellar component to the X-ray emitting gas is
0.2 $h^{-1}$ for the Coma cluster. Assuming Coma is a typical rich
cluster, we can assume that all other clusters have
the same ratio. However, Hattori (1997) suggested that
AXJ2019+1127 at a redshift of 0.94 is a dark cluster
with mass to light ratio $M/L$ $\sim$ 1500 $h$ $M_{\sun}/L_{\sun}$.
Therefore, the stellar component may not be constant from
one cluster to another, but, the effect would be to decrease
the baryonic fraction at low redshifts from the high redshift
values, contrary to what is observed in the present data.
Also, various forms of baryonic components are expected
to exist in the intercluster medium (e.g., 
intercluster stars, Uson {\it et al.} 1991), but, such
 contribution is expected to be lower than the
observed variation in the gas fraction.

\section{Conclusions}
Under the assumption that the baryonic (gas)
fraction in galaxy clusters are constant, we
have constrained the cosmological parameters using the angular diameter
distance relation with redshift. 
However, the present X-ray based data from various
studies may be providing a biased result due to unknown selection
effects. When the angular diameter distance
dependence on the traditional Hubble constant measurements using
SZ and X-ray data are included with the present method
involving cluster gas mass fraction, it is likely that
tighter constraints on the cosmological parameters
are possible. We hope to explore this possibility
in a future paper.
Given the biases that may go in to
the presented gas mass fraction vs. redshift diagram, 
we suggest that results from a well defined and random sample of galaxy
clusters, such as that would be available from AXAF in X-ray 
and from interferometric
observations in SZ,
be used to constrain the cosmological parameters using the
angular diameter distance relation with redshift. 

\section{Note}

Since this work was first submitted, a paper by Danos \& Pen (1998)
appeared, arriving at qualitatively similar conclusions - 
though using cluster electron temperature,
luminosity and angular size to measure the gas mass fraction
of three high redshift clusters, which were compared to a global gas mass
fraction value determined from numerical simulations. 

\begin{acknowledgements}
I would like to thank the  anonymous referee for his/her prompt
refereeing of the paper, and John Carlstrom and Bill Holzapfel
for useful discussions.

\end{acknowledgements}

\clearpage

%\figcaption[fig1.ps]{
%Gas Mass fractions with redshift.
%\label{fig1}}

%\begin{figure}
%\plotone{figure1.eps}
%\end{figure}

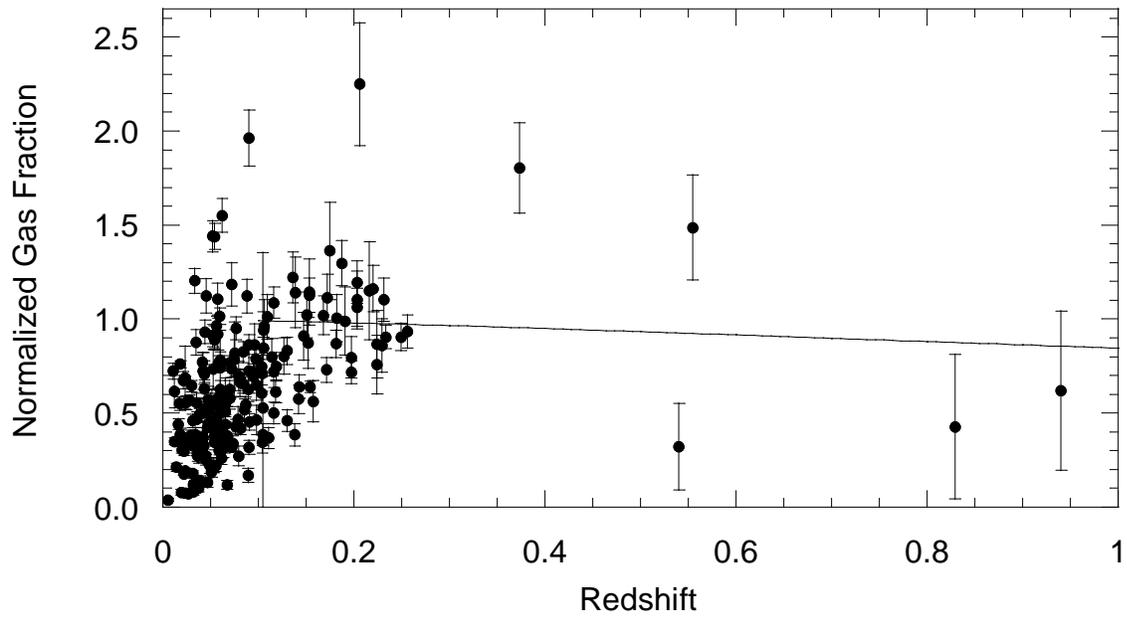
\end{document}